%% file: sample-sigconf.tex
\documentclass[sigconf]{acmart}
\usepackage{subcaption}
\usepackage{multirow}
\usepackage{enumitem}
\usepackage{colortbl}
\definecolor{lightred}{RGB}{245,220,220}
\definecolor{lightgray}{RGB}{245,245,245} 
\AtBeginDocument{%
  }

\copyrightyear{2026}
\acmYear{2026}
\setcopyright{cc}
\setcctype{by-nc-nd}
\acmConference[CIKM '26]{Proceedings of the 35th ACM International Conference on Information and Knowledge Management}{November 07--11, 2026}{Rome, Italy}
\acmBooktitle{Proceedings of the 35th ACM International Conference on Information and Knowledge Management (CIKM '26), November 07--11, 2026, Rome, Italy}
\acmDOI{10.1145/3799682.3840844}
\acmISBN{979-8-4007-2539-5/2026/11}
\begin{document}

\title{Think-to-Personalize: Unifying Reasoning and Retrieval for User-Centric Personalized Dense Retrieval}

\author{Angqing Jiang}
\email{philipgaq@mail.ustc.edu.cn}
\orcid{0009-0000-7897-0742}
\affiliation{
  \institution{University of Science and Technology of China}
  \city{Hefei}
  \state{Anhui}
  \country{China}
}

\author{Gaoming Zhang}
\email{zzzgm@mail.ustc.edu.cn}
\orcid{0009-0006-3179-5602}
\affiliation{
  \institution{University of Science and Technology of China}
  \city{Hefei}
  \state{Anhui}
  \country{China}
}

\author{Jianchun Song}
\email{songjianchun@meituan.com}
\affiliation{
  \institution{Meituan}
  \city{Beijing}
  \country{China}
}

\author{Kena Qi}
\email{qikena@meituan.com}
\affiliation{
  \institution{Meituan}
  \city{Beijing}
  \country{China}
}

\author{Dayao Chen}
\email{chendayao@meituan.com}
\affiliation{
  \institution{Meituan}
  \city{Beijing}
  \country{China}
}

\author{Wei Lin}
\email{linwei31@meituan.com}
\affiliation{
  \institution{Meituan}
  \city{Beijing}
  \country{China}
}

\author{Defu Lian}
\authornote{Corresponding author.}
\email{liandefu@ustc.edu.cn}
\orcid{0000-0002-3507-9607}
\affiliation{
  \institution{University of Science and Technology of China}
  \city{Hefei}
  \state{Anhui}
  \country{China}
}

\renewcommand{\shortauthors}{Jiang et al.}
\newcommand{\name}{Think-to-Personalize}
\newcommand{\nameshort}{TTP}
\begin{abstract}
Dense retrieval has become a cornerstone of modern local-lifestyle e-commerce search by encoding queries and items into semantic embedding spaces. While recent advancements have transitioned from BERT-based embedding models to Large Language Models (LLMs), most approaches still treat LLMs as static text encoders, neglecting their inherent reasoning capabilities. Furthermore, standard dense retrieval models remain query-centric, which is insufficient in e-commerce scenarios where sparse and ambiguous queries create an \textbf{intent gap} that can only be bridged by the rich context of user history.
Meanwhile, existing personalized retrieval methods typically rely on {implicit embedding interactions}, which lack the reasoning capability to effectively disambiguate user intent from noisy historical behaviors.
To address these challenges, we propose \textbf{\name{}} (\textbf{\nameshort{}}), a novel framework that unifies explicit \textbf{user-centric intent reasoning} with dense retrieval. By reasoning over the user's historical purchase sequence, \nameshort{} explicitly deduces latent personalized needs and generates an intent-enhanced query, which is then encoded into a unified dense embedding. Specifically, we design a two-stage training paradigm: (1) A Supervised Fine-Tuning (SFT) stage that establishes cold-start capabilities; (2) A Reinforcement Learning (RL) stage that aligns the reasoning process with retrieval utility using Group Relative Policy Optimization (GRPO). Extensive experiments on both proprietary and public benchmarks demonstrate that \nameshort{} significantly outperforms state-of-the-art baselines. Furthermore, in online A/B tests, it achieved a +0.46\% lift in order volume, validating its practical effectiveness and establishing a new paradigm for reasoning-driven personalized dense retrieval.
\end{abstract}




\maketitle

\input{sections/intro.tex}

\input{sections/related_work.tex}

\input{sections/method.tex}

\input{sections/experiments.tex}

\input{sections/conclusion.tex}

\begin{acks}
This work was supported by the Meituan Research Fund, the Scientific Research Innovation Capability Support Project for Young Faculty, the National Natural Science Foundation of China (NSFC, Grant U24A20253), and the New Generation Artificial Intelligence-National Science and Technology Major Project (Grant 2025ZD0123403).
\end{acks}



\section*{GenAI Usage Disclosure}

Generative AI tools were used in this work under the supervision of the authors. Qwen was used in training data construction to generate intent-enhanced queries. ChatGPT was used for limited writing assistance, including grammar refinement, and limited coding support, including code debugging. The authors are responsible for all scientific decisions, result verification, and the final content of the paper.

\bibliographystyle{ACM-Reference-Format}
\bibliography{sample-base}


\end{document}

%% file: sections/intro.tex
\section{Introduction}

In modern e-commerce search systems, dense retrieval~\cite{guo2016deep, mitra2017neural, guo2020deep} has become an indispensable component for efficiently retrieving relevant items from a large corpus. By encoding queries and items into semantic vector spaces, dense retrieval models enable the system to recall items from a large-scale item pool via Approximate Nearest Neighbor (ANN) search. This paradigm has significantly improved retrieval relevance by capturing deep semantic relationships beyond simple keyword matching. 

With the rapid advancement of Large Language Models (LLMs), the landscape of text embedding models is evolving from traditional encoder-only architectures, such as Contriever~\cite{karpukhin2020dense}, Sentence-BERT~\cite{reimers2019sentence}, BGE~\cite{chen2024m3} to LLM-based embedding models including NV-Embed~\cite{lee2024nv}, Llama2vec~\cite{li2024llama2vec}, and Qwen3-Embedding~\cite{zhang2025qwen3}. Despite their impressive performance, most of them treat LLMs primarily as static text encoders, overlooking their powerful chain-of-thought~\cite{wei2022chain} reasoning capabilities. Recent works such as Search-R3~\cite{gui2025search} and LREM~\cite{tang2025large} have attempted to harness the generative power of LLMs by explicitly generating reasoning chains or query expansions to augment query semantics, then produce a reasoning-augmented query embedding for retrieval. Despite these advancements, the current dense retrieval paradigm remains \textit{query-centric}, ignoring the rich context provided by user history. 
This \textit{query-centric} paradigm is particularly limited in local-lifestyle e-commerce scenarios. In such scenarios, user queries are often sparse and ambiguous, creating an \textit{\textbf{intent gap}} where the query often fails to fully convey the user's latent personalized intent, which is often revealed by the user's history.

While personalized search methods~\cite{ai2017learning,dai2023contrastive} such as UIA~\cite{zeng2023personalized} and ZAM~\cite{ai2019zero} utilize user history for personalized retrieval, they typically rely on implicit user sequence modeling and embedding interactions~\cite{bi2020transformer,zhang2020towards}, lacking the reasoning capability to explicitly deduce latent intent. Consequently, they often fail to capture the complex dependencies between the information-rich historical sequence and the current query's ambiguous intent. To overcome this, recent generative approaches like PBR~\cite{zhang2025personalize} and CoPS~\cite{zhou2024cognitive} have attempted to leverage LLMs to explicitly rewrite queries or generate pseudo-labels based on user history. However, these methods generally operate in a disjoint pipeline, where the LLM is optimized for causal language modeling rather than retrieval utility, leading to a misalignment between the generated rationale and the downstream retrieval objective. 
Thus, a critical challenge remains: how to unify explicit intent reasoning with dense retrieval in an end-to-end manner, thereby bridging the \textit{intent gap} between the literal query and the user's latent personalized intent.

To bridge this intent gap between query semantics and latent personalized intent, we propose \textbf{\name{}} (\textbf{\nameshort{}}), an end-to-end framework that shifts the paradigm from implicit matching to \textit{explicit user-centric intent reasoning}. Unlike conventional approaches, \nameshort{} integrates personalized reasoning and dense retrieval into a unified model. Specifically, it reasons over the user's historical order sequence and current query to deduce latent personalized needs, generating an \textbf{intent-enhanced query}, which is subsequently encoded into an intent-enhanced embedding. This unified framework effectively bridges the intent gap by explicitly uncovering latent needs, while ensuring that the reasoning process is fully aligned with the retrieval objective.

\begin{figure*}[t]
    \centering
    \includegraphics[width=0.7\textwidth]{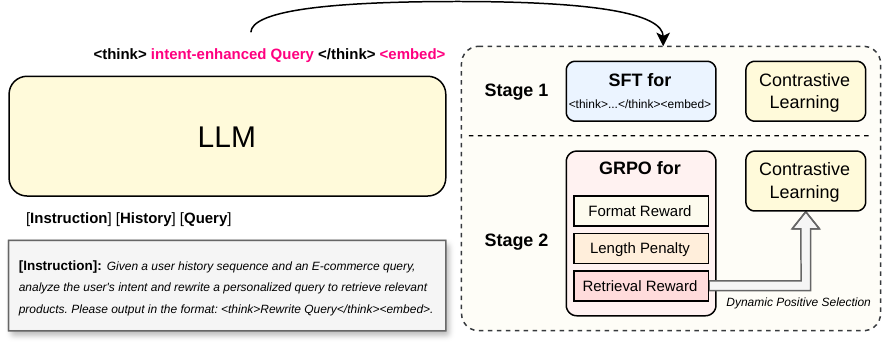}
    \caption{Overview of the \name{} training framework. We first establish cold-start capabilities via SFT with a joint loss. Subsequently, we employ GRPO with a retrieval-aware reward function and dynamic positive selection to align the generated reasoning process with downstream retrieval performance.}
    \Description{The figure depicts the two-stage training paradigm of the framework. The input consists of an instruction, user history, and query. The LLM generates a thinking process (wrapped in 'think' tags) followed by an embedding token. Stage 1 involves Supervised Fine-Tuning (SFT) with a joint generation and contrastive learning objective. Stage 2 introduces Reinforcement Learning using Group Relative Policy Optimization (GRPO). This stage incorporates three reward signals: Format Reward, Length Penalty, and Retrieval Reward. A Dynamic Positive Selection mechanism feeds the best-performing generated queries into the contrastive learning loss, closing the feedback loop.}
    \label{fig:training}
\end{figure*}

To effectively train this framework, we devise a robust two-stage training paradigm as shown in Figure \ref{fig:training}. The first stage employs Supervised Fine-Tuning (SFT) with an InfoNCE loss to establish foundational retrieval capabilities and cold-start performance. The second stage introduces Reinforcement Learning (RL) to further align the quality of the generated intent rationale with downstream retrieval accuracy. Extensive experiments on both proprietary and public benchmarks demonstrate that \nameshort{} significantly outperforms strong general and personalized retrieval baselines, particularly in challenging scenarios involving broad and long-tail queries.
Our code is available at \textbf{\url{https://github.com/PhilipGAQ/ttp}}.

Our contributions are as follows:
\begin{itemize}
    \item We propose \nameshort{}, a novel reasoning-driven framework that unifies user-centric intent reasoning with dense retrieval to produce dynamic, personalized embeddings.
    \item We introduce a novel two-stage training strategy combining SFT and RL, utilizing Group Relative Policy Optimization (GRPO), to effectively align the model's reasoning capabilities with retrieval objectives.
    \item Comprehensive offline and online experiments validate the effectiveness and practical value of our approach in real-world e-commerce scenarios.
\end{itemize}

%% file: sections/related_work.tex
\section{Related Work}
\label{sec:related_work}

\subsection{Dense Retrieval}
\label{subsec:dense_retrieval}
Dense retrieval has evolved from BERT-based dual-encoder architectures, such as DPR~\cite{karpukhin2020dense}, Sentence-BERT~\cite{reimers2019sentence}, and BGE~\cite{chen2024m3}, to LLM-based embedding models and efficient asymmetric encoder designs~\cite{jiang-etal-2026-benchmarking}. State-of-the-art models like Llama2vec~\cite{li2024llama2vec} and NV-Embed~\cite{lee2024nv} leverage the vast semantic knowledge of LLMs to generate high-quality text embeddings using contrastive learning. However, these methods typically utilize LLMs as text encoders, ignoring their inherent generative and reasoning capabilities.
Recently, inspired by the Chain-of-Thought (CoT) capability~\cite{wei2022chain} of LLMs, researchers have begun to integrate reasoning into dense retrieval. O1-Embedder~\cite{yan2025o1} generates thoughts for the query before retrieval. Search-R3~\cite{gui2025search} and LREM~\cite{tang2025large} further advance this by employing Reinforcement Learning to encourage the model to generate explicit reasoning steps before producing embedding, thereby integrating the thinking process into representation learning.
These methods demonstrate that explicit reasoning can significantly clarify ambiguous queries. However, they remain predominantly \textit{query-centric}, focusing on reasoning about the query's surface form (e.g., query expansion or rewriting). In the context of e-commerce, where intent is often deeply personalized, purely query-centric reasoning fails to address the information scarcity of broad and long-tail queries. Our work differs by extending the reasoning scope from the query alone to the user-query pair, enabling the model to bridge the intent gap through user-centric deduction.

\subsection{Personalized Search}
\label{subsec:personalized_search}
Personalized search requires integrating user historical behaviors to disambiguate intent. Prior methods like UIA~\cite{zeng2023personalized} employ a cross-attention mechanism to dynamically aggregate user history representations based on the current query. UEPPR~\cite{jha2024unified} unifies short-term behaviors (queries, clicks) and long-term preferences (purchase tags, geographical preferences) into a Query-User Tower, utilizing lightweight Transformers for in-session attention. OneSearch~\cite{chen2025onesearch} introduces an end-to-end generative retrieval framework that incorporates short-term behaviors into prompts and compresses long-term preferences into RQ-OPQ vectors, directly generating item IDs under relevance constraints. However, these approaches rely on implicit modeling of user behavior sequences, lacking explicit reasoning capabilities for complex intent understanding.
With the development of LLMs, recent works have explored generative query enhancement for personalization~\cite{chen2023graph, nguyen2025rl}. CoPS~\cite{zhou2024cognitive} integrates LLMs with a memory mechanism for personalized search experiences. PBR~\cite{zhang2025personalize} leverages LLMs to generate pseudo-queries and reasoning chains based on user history. However, these methods typically operate in a \textit{disjoint pipeline}: the LLM acts as an external re-writer, optimized separately from the downstream retriever. This separation prevents the reasoning process from being directly aligned with the final retrieval objective.
Recent LLM-based studies have also explored adjacent issues, including adaptive semantic capacity, model reliability, and long-horizon memory management~\cite{xu2026lever,liu2026dualpathwaycircuitsobjecthallucination,liu2026conmemcontributionawarememorylonghorizon,zhang2026memmarkstateevolutionattributionwatermarking}.
In contrast, our work unifies the reasoning and retrieval processes into an end-to-end framework, where the reasoning process is explicitly optimized to maximize downstream retrieval utility.

\subsection{Reinforcement Learning for LLM}
\label{subsec:reinforcement_learning}
Reinforcement Learning (RL) has become a critical component in LLM training pipelines, enhancing model controllability and performance~\cite{wang2025reinforcementlearningenhancedllms}. For instance, Reinforcement Learning from Human Feedback (RLHF)~\cite{ouyang2022training}, typically employs Proximal Policy Optimization (PPO)~\cite{schulman2017proximal} to align models with human preferences. PPO optimizes a policy model using scalar signals from a reward model while constraining deviations via a reference model and estimating future value with a critic network. Despite its success in models like InstructGPT~\cite{ouyang2022training}, PPO suffers from high computational complexity and training instability due to its multi-model interaction. To address this, Direct Preference Optimization (DPO)~\cite{rafailov2023direct} simplifies alignment by modeling preference learning as a classification problem, eliminating the need for an explicit reward model. While efficient, DPO may struggle with complex reasoning tasks where intermediate steps are crucial. Recently, Group Relative Policy Optimization (GRPO)~\cite{shao2024deepseekmath}, introduced in DeepSeekMath, offers a compelling alternative. By replacing the value model with a group-sampling-based baseline estimation, GRPO significantly reduces computational overhead while maintaining the optimization stability required for mathematical and logical reasoning. 

%% file: sections/method.tex
\section{Methods}
\label{sec:method}

\begin{table}[t]
    \centering
    \caption{The instruction prompt used for intent reasoning.}
    \label{tab:instruction}
    \small
    \begin{tabular}{|p{0.95\columnwidth}|}
    \hline
    \rowcolor{lightred} \textbf{Instruction Template} \\
    \hline
    \rowcolor{lightgray} {Given a user history sequence and an E-commerce query, analyze the user's intent and rewrite a personalized query to retrieve relevant products. Please output in the format: <think>Rewrite Query</think><embed>} \\
    \hline
    \end{tabular}
\end{table}

\subsection{Framework Overview}
We consider open-text personalized dense retrieval in e-commerce. Let $\mathcal{U}$ be the user set and $\mathcal{I}$ the item set. For each user $u$, we observe a historical behavior sequence $H_u = \{i_{u,1}, \dots, i_{u,t}\}$ consisting of past purchased items, where each item is represented by its textual metadata (e.g., title, category, shop name; see Section~\ref{subsec:implementation_details} for details). Given a query $q$, the goal is to retrieve the most relevant target item $i^+ \in \mathcal{I}$.

Figure~\ref{fig:training} illustrates the overall framework of \nameshort{}. Unlike traditional query-centric dense retrieval methods, which encode the current query alone, \nameshort{} explicitly incorporates both user history and intent reasoning into the query encoding process. The query-side input is constructed as
\begin{equation}
    x = \texttt{[INST]} \oplus I_{\text{task}} \oplus H_u \oplus q \oplus \texttt{[/INST]},
\end{equation}
where $I_{\text{task}}$ is the instruction (see Table~\ref{tab:instruction}). The model then generates an intent-enhanced query $r$ enclosed in \texttt{<think>} and \texttt{</think>} tokens, followed by a special \texttt{<embed>} token:
\begin{equation}
    y = \texttt{<think>} \oplus r \oplus \texttt{</think>} \oplus \texttt{<embed>}
\end{equation}
The final query embedding $\mathbf{e}_q$ is extracted from the last hidden state of \texttt{<embed>}. On the item side, we construct the item input by serializing its textual metadata and appending the same \texttt{<embed>} token, and extract the item embedding $\mathbf{e}_i$ from the last hidden state of this token using the same shared encoder. The relevance score is computed by $s(q, i) = \mathbf{e}_q^\top \mathbf{e}_i$.

To train this unified framework, we adopt a two-stage strategy. SFT first initializes the model to follow the reasoning-and-retrieval format while learning retrieval-oriented representations. RL then further aligns the reasoning process with downstream retrieval utility using GRPO, a retrieval-aware reward, and dynamic positive selection. Since the query and item towers share the same encoder, both sides are updated symmetrically throughout training. This design allows retrieval supervision to directly shape the generated personalized intent.

\begin{figure*}[t]
    \centering
    \includegraphics[width=0.7\textwidth]{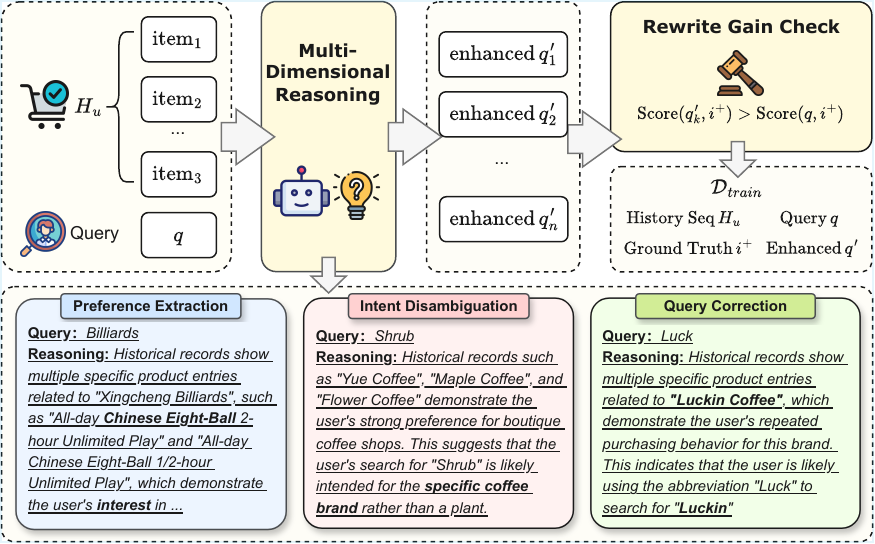}
    \caption{Overview of the reasoning-enhanced data construction pipeline. A large teacher model (Qwen3-32B) performs chain-of-thought reasoning to generate intent-enhanced queries, which are then filtered by a re-ranker to ensure high retrieval utility.}
    \Description{The figure illustrates the data construction process of the proposed framework. It shows a pipeline starting with user history and a query, passing through a multi-dimensional reasoning module (represented by a robot icon), and outputting enhanced queries. A 'Rewrite Gain Check' step filters these queries before training. Below, three specific examples demonstrate the reasoning types: (1) Preference Extraction: inferring 'Billiards' interest from 'Chinese Eight-Ball' history; (2) Intent Disambiguation: clarifying 'Shrub' as specific coffee brands like 'Yue Coffee' based on coffee-related history; (3) Query Correction: mapping 'Luck' to 'Luckin Coffee' based on purchase history.}
    \label{fig:data_construction}
\end{figure*}

\subsection{Data Construction}

\begin{table}[t]
    \centering
    \caption{The teacher prompt template used for intent reasoning.}
    \label{tab:teacher_prompt}
    \scriptsize
    \setlength{\tabcolsep}{4pt}
    \begin{tabular}{|p{0.95\columnwidth}|}
    \hline
    \rowcolor{lightred} \textbf{Teacher Prompt} \\
    \hline
    \rowcolor{lightgray}
    You are a personalized query rewriting assistant for e-commerce retrieval. \newline
    Given a user's historical behavior sequence and the current query, first reason over the history and query, and then generate multiple candidate rewritten queries for retrieval. \newline
    Consider the following four cases: \newline
    \textbf{1. Preference Extraction:} If the current query already conveys a relatively clear intent, use the user history to extract query-relevant preferences and refine the query into a more specific and personalized form, while preserving its original intent. \newline
    \textbf{2. Intent Disambiguation:} If the current query is broad, ambiguous, or underspecified, use the user history to infer the most likely intended meaning and rewrite the query toward that personalized intent. \newline
    \textbf{3. Query Correction:} If the current query is noisy, incomplete, misspelled, abbreviated, or otherwise difficult to retrieve effectively, use the user history to correct or complete it into a clearer and more standard form. \newline
    \textbf{4. No Rewrite:} If the user history does not provide clear and relevant evidence, or if personalization may distort the original intent, keep the original query unchanged. \newline
    \textbf{Requirements:} First provide a brief reasoning process grounded in the history and query; use only query-relevant evidence from the user history; preserve the core meaning of the original query unless correction is necessary; do not introduce irrelevant or conflicting intents; keep each rewrite concise and retrieval-oriented. \newline
    User History: \{history\} \newline
    Query: \{query\} \newline
    Output: \{"reasoning": "...", "rewritten\_queries": ["...", "...", "...", "...", "..."]\} \\
    \hline
    \end{tabular}
\end{table}

High-quality reasoning data is crucial for training. We first filter the user history context $H_u$ based on the user's purchase records from the log data over the past year. For each item, we serialize it into a text string by concatenating its metadata fields (e.g., Title, Category, Menu). To manage sequence length while retaining the most informative context, for users with more than $N$ historical items, we employ the open-source \textit{bge-reranker-v2-m3}~\cite{chen2024m3} re-ranker to select the top-$N$ items most relevant to the current query $q$. The selected items are then re-ordered chronologically and used as the final truncated history sequence for teacher generation and downstream model training.

We employ Qwen3-32B~\cite{yang2025qwen3} to construct a reasoning-enhanced dataset. As shown in Figure \ref{fig:data_construction}, given the instruction $I_{task}$, the processed history $H_u$, and query $q$, we prompt the teacher model to explicitly perform Chain-of-Thought (CoT) reasoning to infer the user's personalized search intent before generating intent-enhanced query candidates. The full prompt template is shown in Table~\ref{tab:teacher_prompt}. Specifically, the model is instructed to perform multi-dimensional reasoning over the user history and query, focusing on three key aspects: (1) \textbf{Preference Extraction}: identifying latent personalized preference relevant to the current search (e.g., brand affinity, stylistic vibe); (2) \textbf{Intent Disambiguation}: resolving broad or underspecified queries by grounding them in the user's behavioral context; and (3) \textbf{Query Correction}: implicitly correcting potential typos, incomplete inputs, or vague descriptions based on historical context to recover the precise search intent. Based on this comprehensive analysis, the teacher model generates multiple candidate \textbf{intent-enhanced queries} $q'$. To minimize inference latency for the student model, we discard the intermediate reasoning steps and only retain the enhanced query as the training target. Furthermore, to prevent semantic drift, we explicitly instruct the teacher to preserve the original query $q$ if the user's history provides no relevant context, thereby avoiding over-personalization.

To ensure quality, we generate $K=5$ candidate rewrites $\{q'_1, ..., q'_K\}$ for each sample using the default decoding configuration of Qwen3-32B. We then use the open-source \textit{bge-reranker-v2-m3} re-ranker to score the relevance of each rewrite against the ground-truth positive item $i^+$, which is the user's final purchased item of the current query. After filtering out rewrites with non-positive relevance gain, we uniformly sample one training target from the top-3 remaining candidates ranked by gain. To prevent distributional bias towards over-rewriting, we also retain a proportion of neutral samples where the teacher decides to keep the original query, ensuring the model learns when to preserve user input.

\subsection{Cold-Start SFT}

In the first stage, we use Supervised Fine-Tuning (SFT) to establish the model's capability to strictly follow the input format and generate valid embeddings. We add \texttt{<think>}, \texttt{</think>}, and \texttt{<embed>} to the tokenizer's special tokens. The training data consists of pairs $(x, \hat{y})$, where $x$ is the instruction-augmented input and $\hat{y}$ is the high-quality target sequence selected from the data construction step.

We optimize the model using a joint loss combining generation and contrastive learning. First, $\mathcal{L}_{gen}$ is the standard Next Token Prediction (NTP) loss applied to the reasoning content and special tokens:
\begin{equation}
    \mathcal{L}_{\text{gen}} = - \frac{1}{|\hat{y}|} \sum_{t=1}^{|\hat{y}|} \log P(\hat{y}_t | \hat{y}_{<t}, x)
\end{equation}
Second, for the contrastive objective, we extract the query embedding $\mathbf{e}_q$ from the last hidden state of the \texttt{<embed>} token. The InfoNCE loss is formulated as:
\begin{equation}
    \mathcal{L}_{\text{cl}} = -\log \frac{\exp(s(q, i^+) / \tau)}{\exp(s(q, i^+) / \tau) + \sum_{j \in \mathcal{N}} \exp(s(q, i^-_j) / \tau)}
\end{equation}
where $\mathcal{N}$ is the set of in-batch negatives. We utilize in-batch negatives for efficient training~\cite{karpukhin2020dense}. Finally, the total Stage 1 loss is formulated as:
\begin{equation}
    \mathcal{L}_{\text{Stage1}} = \lambda_{\text{gen}} \mathcal{L}_{\text{gen}} + \mathcal{L}_{\text{cl}}
\end{equation}
In our experiments, we set the generation weight $\lambda_{\text{gen}} = 0.5$ to balance the structural learning of the reasoning format with the semantic alignment for retrieval. This stage serves as a cold start, enabling the model to effectively initialize the generation and retrieval capabilities.

\subsection{Reinforcement Learning Alignment}
While SFT provides a good initialization, the generated reasoning may not always be optimal for retrieval. In Stage 2, we employ Group Relative Policy Optimization (GRPO) combined with InfoNCE to align the reasoning generation with retrieval performance.

\subsubsection{Reward Modeling}
We design a composite reward function $R = \lambda_{\text{fmt}} R_{\text{format}} + \lambda_{\text{len}} R_{\text{length}} + \lambda_{\text{ret}} R_{\text{retrieval}}$ to guide the policy optimization.

\textbf{Format Reward} ($R_{\text{format}}$): This component enforces structural validity.
\begin{equation}
    R_{\text{format}} = \begin{cases} 
    0 & \text{if output follows the \texttt{<think>} format} \\
    -1 & \text{otherwise}
    \end{cases}
\end{equation}

\textbf{Length Penalty} ($R_{\text{length}}$): To prevent verbosity and reduce latency, we penalize outputs exceeding a length threshold $L$:
\begin{equation}
    R_{\text{length}} = - \max(0, |y| - L)
\end{equation}

\textbf{Retrieval Reward} ($R_{\text{retrieval}}$): We leverage the frozen SFT model as both the reference model for KL divergence calculation and the retrieval reward model. Given that the SFT model has established strong discriminative capabilities for retrieval, it serves as a stable semantic anchor which is intrinsically aligned with the final retrieval objective. By freezing this model, we ensure that policy optimization is guided by stable and reliable retrieval signals rather than an evolving reward space, thereby preventing self-referential bias and reward drift while still allowing retrieval to guide the reasoning generation. For a generated intent-enhanced query $q'$, we calculate the retrieval reward based on the relevance gain:
\begin{equation}
    R_{\text{retrieval}} = \alpha \cdot \Delta S_{\text{pos}} + \beta \cdot \Delta S_{\text{margin}}
\end{equation}
where $\Delta S_{\text{pos}} = s(q', i^+) - s(q, i^+)$ and $\Delta S_{\text{margin}}$ is the margin improvement against the average score of negative samples $\mathcal{N}$:
\begin{equation}
    \Delta S_{\text{margin}} = {\left( s(q', i^+) - \bar{S}(q', \mathcal{N}) \right)} - {\left( s(q, i^+) - \bar{S}(q, \mathcal{N}) \right)}
\end{equation}
The $\Delta S_{\text{pos}}$ term directly encourages the model to generate queries with higher relevance to the target item, while the $\Delta S_{\text{margin}}$ term encourages the model to enhance the discriminative power over the negative samples. To stabilize training, we clip the final retrieval reward $R_{\text{retrieval}}$ to the range $[-1, 1]$.

\subsubsection{Optimization Objective}
We formulate the final objective as a multi-task learning problem combining policy optimization and representation learning.

\textbf{GRPO Objective}: For each input $x$, we sample a group of $G$ outputs $\{y_1, ..., y_G\}$ from the old policy $\pi_{\theta_{old}}$. Let $\rho_i = \frac{\pi_\theta(y_i|x)}{\pi_{\theta_{old}}(y_i|x)}$ be the probability ratio. GRPO optimizes the policy by maximizing the advantage of high-reward outputs without a value network:
\begin{equation}
\begin{aligned}
    \mathcal{L}_{\text{GRPO}} = - \frac{1}{G} \sum_{i=1}^G \bigg( & \min \left( \rho_i A_i, \text{clip}\left(\rho_i, 1-\epsilon, 1+\epsilon\right) A_i \right) \\
    & - \beta_{\text{KL}} D_{\text{KL}}(\pi_\theta(y_i|x) || \pi_{\text{ref}}(y_i|x)) \bigg)
\end{aligned}
\end{equation}
where $A_i = \frac{R_i - \bar{R}}{\sigma_R}$ represents the normalized advantage.

\textbf{InfoNCE with Dynamic Positive Selection}: To ensure robust representation learning, we introduce a \textbf{Dynamic Positive Selection} strategy. Instead of using all sampled outputs for contrastive learning, we select the best rationale $y^*$ from the group:
\begin{equation}
    y^* = \operatorname*{argmax}_{y_i \in \{y_1...y_G\}} R_{\text{retrieval}}(y_i)
\end{equation}
We then conditionally update the embedding space. If $R_{\text{retrieval}}(y^*) > 0$, we use the embedding of $y^*$ as the positive query representation; otherwise, we fallback to the embedding of the original query $q$. This effectively denoises the contrastive learning signal. The InfoNCE loss $\mathcal{L}_{\text{cl}}$ is calculated using this selected embedding.

The total objective for Stage 2 is:
\begin{equation}
    \mathcal{L}_{\text{Stage2}} = \mathcal{L}_{\text{GRPO}} + \lambda_{\text{cl}} \mathcal{L}_{\text{cl}}
\end{equation}

%% file: sections/experiments.tex
\section{Experiments}
\label{sec:experiments}

\begin{table}[t]
    \centering
    \caption{Statistics of the evaluation datasets.}
    \label{tab:dataset_stats}
    \resizebox{0.95\columnwidth}{!}{
    \begin{tabular}{l|c|c|c|c}
    \toprule
    \textbf{Dataset} & \textbf{\# Queries} & \textbf{\# Items} & \textbf{Avg. Hist} & \textbf{Avg. Q-Len} \\
    \midrule
    General & 15,795 & \multirow{3}{*}{4,481,107} & 8.1 & 7.3 \\
    Broad & 14,634 & & 8.1 & 6.2 \\
    LongTail & 7,746 & & 7.8 & 10.1 \\
    \midrule
    KuaiSearch & 8,468 & 6,634,118 & 16.5 & 10.0 \\
    Amazon & 2,174 & 35,772 & 40.1 & 36.0 \\
    \midrule
    Relevance & 300,000 & 300,000 & - & 7.4 \\
    \bottomrule
    \end{tabular}
    }
\end{table}

\subsection{Experimental Setup}

\subsubsection{Datasets and Evaluation Metrics}
We evaluate \nameshort{} on both proprietary and public benchmarks. For proprietary benchmarks, we construct four test sets derived from real-world logs with different query characteristics, as summarized in Table~\ref{tab:dataset_stats}. Specifically, \textbf{General} contains over 15,000 user queries sampled from online search logs, each paired with the user's historical purchase sequence and the actually purchased item as ground truth. \textbf{Broad} focuses on ambiguous or underspecified queries and is designed to evaluate intent disambiguation. \textbf{LongTail} contains low-frequency queries sampled from the tail 10\% of the real-world query distribution. \textbf{Relevance} is a history-free relevance test set consisting of 300,000 query-item pairs annotated by an internal domain-adaptive 13B relevance LLM with three relevance levels.
We further evaluate on two public benchmarks, \textbf{KuaiSearch~\cite{li2026kuaisearch}} and \textbf{Amazon}.\footnote{Although public datasets such as JDSearch~\cite{liu2023jdsearch} and KuaiSAR~\cite{sun2023kuaisar} exist, they only provide encrypted or ID-based token texts rather than natural-language item content, making them unsuitable for evaluating intent reasoning in our setting.} For KuaiSearch, we use the \textit{lite} recall version with user history constructed from both historical clicks and purchases; for Amazon, we use the widely adopted PersonalWAB~\cite{cai2025large, qin2025maps} synthetic-query version constructed from Amazon Reviews~\cite{kang2018self}, where user history is constructed from past purchase sequences as in our proprietary setting. We follow the original split and evaluation protocol.

For the General, Broad, and LongTail sets, we report {Recall@K} ($K=10,20,100$) and {MRR@100}. For the Relevance set, we report {Spearman Correlation} between model scores and relevance labels. For public benchmarks, we follow their original evaluation protocols and report the corresponding standard metrics.

\subsubsection{Baseline Methods}
We compare \nameshort{} with three categories of baselines. \textbf{(1) General retrievers}: BM25~\cite{robertson2009probabilistic}, BERT-base, and Qwen-Embedding. BERT-base and Qwen-Embedding are dual-encoder dense retrievers built on BERT-base~\cite{devlin2019bert} and Qwen2.5-3B-Instruct~\cite{yang2025qwen3}, respectively, and trained with contrastive learning only. \textbf{(2) Implicit history modeling}: UIA~\cite{zeng2023personalized}, History-Aware, MAPs~\cite{qin2025maps}, and CoPPS~\cite{dai2023contrastive}. History-Aware is a history-concatenation dense retriever baseline, where the serialized user history is prepended to the query using the same history serialization strategy as \nameshort{}, but trained solely with contrastive learning. \textbf{(3) Generative-Enhanced methods}: HyDE~\cite{gao2023precise}, P-PRF~\cite{zhang2025personalize}, and Decoupled-Stage. For all generative baselines, we use Qwen2.5-3B-Instruct as the generator and Qwen-Embedding as the retriever. In particular, Decoupled-Stage is our strongest rewrite-then-retrieve baseline, where the rewriter is fine-tuned on the same SFT data as \nameshort{}, then frozen to generate rewritten queries for training a separate retriever with the same retrieval supervision as our dense retriever.

For fair comparison, our re-implemented baselines use the same training split and aligned optimization settings whenever applicable. Methods marked with * are quoted from the original papers; the remaining methods are re-implemented in our experiments.

\subsubsection{Implementation Details}
\label{subsec:implementation_details}
We adopt Qwen2.5-3B-Instruct~\cite{yang2025qwen3} as the backbone. For our proprietary platform, we first use Qwen3-32B to generate and clean 1M intent-enhanced training samples following the data construction procedure described in Section~\ref{sec:method}. We use the full cleaned dataset for SFT initialization, and then select the top 30\% highest-gain samples ranked by rewrite gain for RL. This design allows RL to focus on cases where rewrite quality most strongly affects retrieval. For KuaiSearch and Amazon, we strictly follow their released benchmark splits and apply the same teacher-data construction, using the same SFT/RL data allocation ratio as in the proprietary setting. The resulting training sets contain 287,093 samples for KuaiSearch and 6,896 samples for Amazon (PersonalWAB).
We serialize each historical item and candidate item into natural-language text by concatenating its metadata fields in the form of \texttt{key:value}, separated by \texttt{[SEP]}. For the proprietary platform, the main fields include title, product, menu, and geolocation; for KuaiSearch, we use brand name, title, category, seller; and for Amazon, we use title, main category, and features.

For the SFT stage, we employ LoRA~\cite{hu2022lora} fine-tuning with $r=16$ and $\alpha=32$. The per-device batch size is set to 64, and the learning rate is $1e-4$. We use the generation loss weight $\lambda_{\text{gen}}=0.5$, and the InfoNCE temperature $\tau=0.02$. Both the query and item encoders share the same LLM backbone and are updated simultaneously, with an embedding dimension of 2048. Query-side and item-side inputs are both truncated to 512 tokens; when the input exceeds this limit, we first truncate the history sequence while preserving the instruction, current query, and core item content. All contrastive learning objectives in our experiments, including SFT, RL, and distillation, use in-batch negatives.
For the RL stage, we implement the RL training based on the VeRL~\cite{sheng2024hybridflow} framework. The group size for GRPO is set to $N=8$. We train for 3 epochs with a learning rate of $2e-7$ and a total training batch size of 128. For the reward configuration, we set the format weight $\lambda_{\text{fmt}}=0.5$, length penalty weight $\lambda_{\text{len}}=0.5$ (with threshold $L=64$ tokens), and retrieval reward weight $\lambda_{\text{ret}}=2.0$. To stabilize training, we calculate $R_{\text{retrieval}}$ using $\alpha=2.0$ and $\beta=1.0$, and clip the final result to $[-1, 1]$. Finally, the InfoNCE loss weight $\lambda_{\text{cl}}$ in the total objective is set to 0.1. All experiments are conducted on 8 NVIDIA A100 GPUs.

\subsection{Offline Evaluation}
\subsubsection{Main Results}

\begin{table*}[t]
    \caption{Overall performance comparison on the four test sets. The best results are highlighted in \textbf{bold}, and the second-best results are \underline{underlined}. \nameshort (SFT Only) and \nameshort (SFT+RL) represent our method trained with only the SFT stage and the full two-stage pipeline, respectively. Improvements of \nameshort (SFT+RL) over the best baseline are statistically significant ($p < 0.05$). Note that the Relevance dataset does not have user history, so baseline methods relying on user history have no results.}
    \label{tab:main_results}
    \resizebox{\textwidth}{!}{
    \begin{tabular}{l|cccc|cccc|cccc|c}
    \toprule
    \multirow{2}{*}{\textbf{Method}} & \multicolumn{4}{c|}{\textbf{General}} & \multicolumn{4}{c|}{\textbf{Broad}} & \multicolumn{4}{c|}{\textbf{LongTail}} & \textbf{Relevance} \\
    \cmidrule(lr){2-5} \cmidrule(lr){6-9} \cmidrule(lr){10-13} \cmidrule(lr){14-14}
     & \textbf{R@10} & \textbf{R@20} & \textbf{R@100} & \textbf{MRR} & \textbf{R@10} & \textbf{R@20} & \textbf{R@100} & \textbf{MRR} & \textbf{R@10} & \textbf{R@20} & \textbf{R@100} & \textbf{MRR} & \textbf{Spearman} \\ \midrule
    \multicolumn{14}{l}{\textit{General Retrievers}} \\
    BM25 & 31.01 & 42.37 & 61.96 & 13.28 & 31.46 & 42.04 & 58.80 & 12.65 & 22.48 & 29.15 & 42.50 & 11.14 & - \\
    BERT-base & 31.77 & 42.83 & 63.82 & 12.84 & 33.01 & 42.65 & 60.39 & 12.83 & 21.86 & 28.36 & 42.89 & 9.52 & 45.70 \\
    Qwen-Embedding & 33.25 & 44.20 & 64.55 & 14.01 & 31.89 & 41.27 & 58.15 & 13.42 & 23.14 & 29.76 & 45.08 & 10.66 & 48.51 \\
    \midrule
    \multicolumn{14}{l}{\textit{Implicit History Modeling}} \\
    UIA & 31.06 & 42.19 & 62.87 & 14.14 & 33.45 & 42.25 & 57.67 & 15.79 & 24.28 & 32.18 & 46.44 & 11.61 & - \\ 
    History-Aware & 33.23 & 42.47 & 61.18 & 15.27 & 33.70 & 42.39 & 58.08 & 15.82 & 25.29 & 32.66 & 46.86 & 12.44 & - \\ 
    MAPs & 33.31 & 45.76 & 66.03 & 14.92 & 34.62 & 45.57 & 61.01 & 16.18 & 25.45 & 33.84 & 47.19 & 13.25 & - \\
    \midrule
    \multicolumn{14}{l}{\textit{Generative-Enhanced Methods}} \\
    HyDE & 27.93 & 36.86 & 53.81 & 12.76 & 24.61 & 31.29 & 44.17 & 11.43 & 23.14 & 29.76 & 45.08 & 10.66 & - \\
    P-PRF & 33.78 & 43.06 & 64.96 & 14.82 & 34.85 & 43.76 & 60.99 & 15.01 & 25.73 & 32.91 & 46.12 & 12.31 & - \\ 
    Decoupled-Stage & 34.54 & \underline{45.89} & \underline{65.03} & 14.76 & 36.14 & 45.08 & 60.70 & 16.21 & 27.57 & 35.18 & 51.13 & 13.41 & - \\
    \midrule
    \multicolumn{14}{l}{\textit{Our Method}} \\
    \textbf{\nameshort (SFT Only)} & \underline{34.64} & {45.38} & {65.01} & \underline{15.34} & \underline{36.43} & \underline{45.66} & \underline{61.06} & \underline{16.33} & \underline{27.96} & \underline{35.70} & \underline{51.67} & \underline{13.89} & \underline{49.43} \\
    \textbf{\nameshort (SFT+RL)} & \textbf{36.64} & \textbf{47.52} & \textbf{68.02} & \textbf{16.78} & \textbf{39.31} & \textbf{48.74} & \textbf{65.12} & \textbf{17.91} & \textbf{30.30} & \textbf{38.65} & \textbf{54.47} & \textbf{15.04} & \textbf{50.09} \\ \bottomrule
    \end{tabular}}
    \end{table*}

Table~\ref{tab:main_results} presents the overall performance comparison of \nameshort{} against baselines across the four test sets. The results highlight several key observations:

\textbf{(1) Importance of Historical Context.} Methods that incorporate user history (Implicit Modeling, Generative-Enhanced, and ours) consistently outperform general retrievers that rely solely on the current query. This performance gap validates the prevalence of the \textit{intent gap} in e-commerce scenarios, confirming that historical behaviors provide essential context for disambiguating sparse user queries.

\textbf{(2) Overall Superiority.} \nameshort{} (SFT+RL) establishes a new state-of-the-art performance, consistently outperforming all competitive baselines across every metric. Specifically, it surpasses the strongest implicit baseline (\textit{MAPs}) by +1.76\% Recall@20 and +1.99\% Recall@100 on the General dataset. This validates that explicitly reasoning with strong LLMs captures latent user intents more effectively than implicit modeling, which often struggles to distill signals from noisy history. Furthermore, \nameshort{} exceeds the stronger decoupled retrieval baseline \textit{Decoupled-Stage} by +1.63\% Recall@20 and +2.99\% Recall@100 on the General dataset, demonstrating the superiority of our unified end-to-end framework over disjoint pipeline approaches that separate reasoning from retrieval.

\textbf{(3) Bridging the Intent Gap.} The advantages of \nameshort{} are most pronounced in challenging scenarios characterized by high ambiguity. On the \textbf{Broad} dataset, \nameshort{} outperforms the base retriever (Qwen-Embedding) by a remarkable margin of +7.47\% in Recall@20, significantly higher than the gain on the General dataset (+3.32\%). On the \textbf{LongTail} dataset, it surpasses the stronger Decoupled-Stage baseline by +3.47\% Recall@20, compared to +1.63\% Recall@20 on the General dataset. These results provide empirical evidence that for sparse and ambiguous queries, the user's latent intent is inherently rooted in their history. Our framework effectively mines this context to explicitly bridge the intent gap, enabling precise retrieval where query-centric models fail.

\textbf{(4) Effectiveness of RL Alignment.} Comparing \nameshort (SFT Only) and \nameshort (SFT+RL), we observe that the RL stage contributes to consistent gains across all metrics. For instance, Recall@10 on the General set improves from 34.64\% to 36.64\%. Crucially, on the \textbf{Relevance} test set, the Spearman correlation increases from 49.43 to 50.09. This confirms that the RL stage not only refines the model's personalized reasoning capabilities but also preserves its fundamental semantic discrimination, thereby ensuring robustness even in cold-start scenarios where historical context is unavailable.

\textbf{(5) Public Benchmark Generalization.} We further validate \nameshort{} on two public benchmarks, Amazon (PersonalWAB synthetic queries) and KuaiSearch, as shown in Table~\ref{tab:public_results}. On Amazon, \nameshort{} (SFT+RL) outperforms both recent personalized baselines, CoPPS and MAPs, as well as the stronger Decoupled-Stage baseline, while RL further improves over the SFT-only variant. On KuaiSearch, \nameshort{} again achieves the best overall results and consistently surpasses both Decoupled-Stage and \nameshort{} (SFT Only). These results show that the gains of \nameshort{} generalize beyond our proprietary platform under different public benchmark settings.

\begin{table}[t]
\centering
\caption{Public benchmark validation on Amazon (PersonalWAB synthetic queries) and KuaiSearch. Methods marked with * are quoted from the
corresponding original papers.}
\label{tab:public_results}
\scriptsize
\setlength{\tabcolsep}{3pt}
\resizebox{0.95\columnwidth}{!}{
\begin{tabular}{l|cc|cc}
\toprule
\multirow{2}{*}{\textbf{Method}} & \multicolumn{2}{c|}{\textbf{Amazon}} & \multicolumn{2}{c}{\textbf{KuaiSearch}} \\
 & \textbf{HR@20} & \textbf{nDCG@20} & \textbf{HR@20} & \textbf{R@20} \\
\midrule
BM25 & 65.04 & 44.90 & 14.27 & 10.37 \\
Qwen-Embedding & 73.21 & 47.19 & 18.81 & 13.73 \\
History-Aware & 78.57 & 48.92 & 19.71 & 14.34 \\
CoPPS* & 72.86 & 34.39 & - & - \\
MAPs* & 89.87 & 49.95 & - & - \\
Decoupled-Stage & 87.17 & 49.93 & 20.04 & 14.27 \\
\nameshort{} (SFT Only) & 87.85 & 50.09 & 20.73 & 14.76 \\
\textbf{\nameshort{} (SFT+RL)} & \textbf{90.17} & \textbf{51.71} & \textbf{21.94} & \textbf{16.21} \\
\bottomrule
\end{tabular}
}
\end{table}

\subsubsection{Qualitative Analysis}

\begin{figure}[t]
    \centering
    \includegraphics[width=0.85\columnwidth]{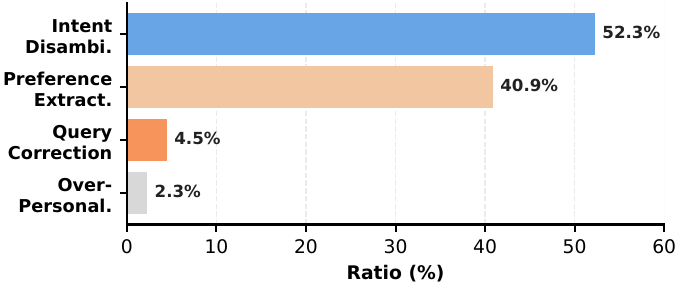}
    \caption{Distribution of rewrite types among rewritten cases only on 200 broad-query examples.}
    \Description{A chart showing the distribution of rewrite types among rewritten broad-query cases only for TTP. Intent Disambiguation accounts for 52.3 percent, Preference Extraction for 40.9 percent, Query Correction for 4.5 percent, and Over-Personalization for 2.3 percent.}
    \label{fig:rewrite_behavior_distribution}
\end{figure}

\begin{figure}[t]
    \centering
    \includegraphics[width=\columnwidth]{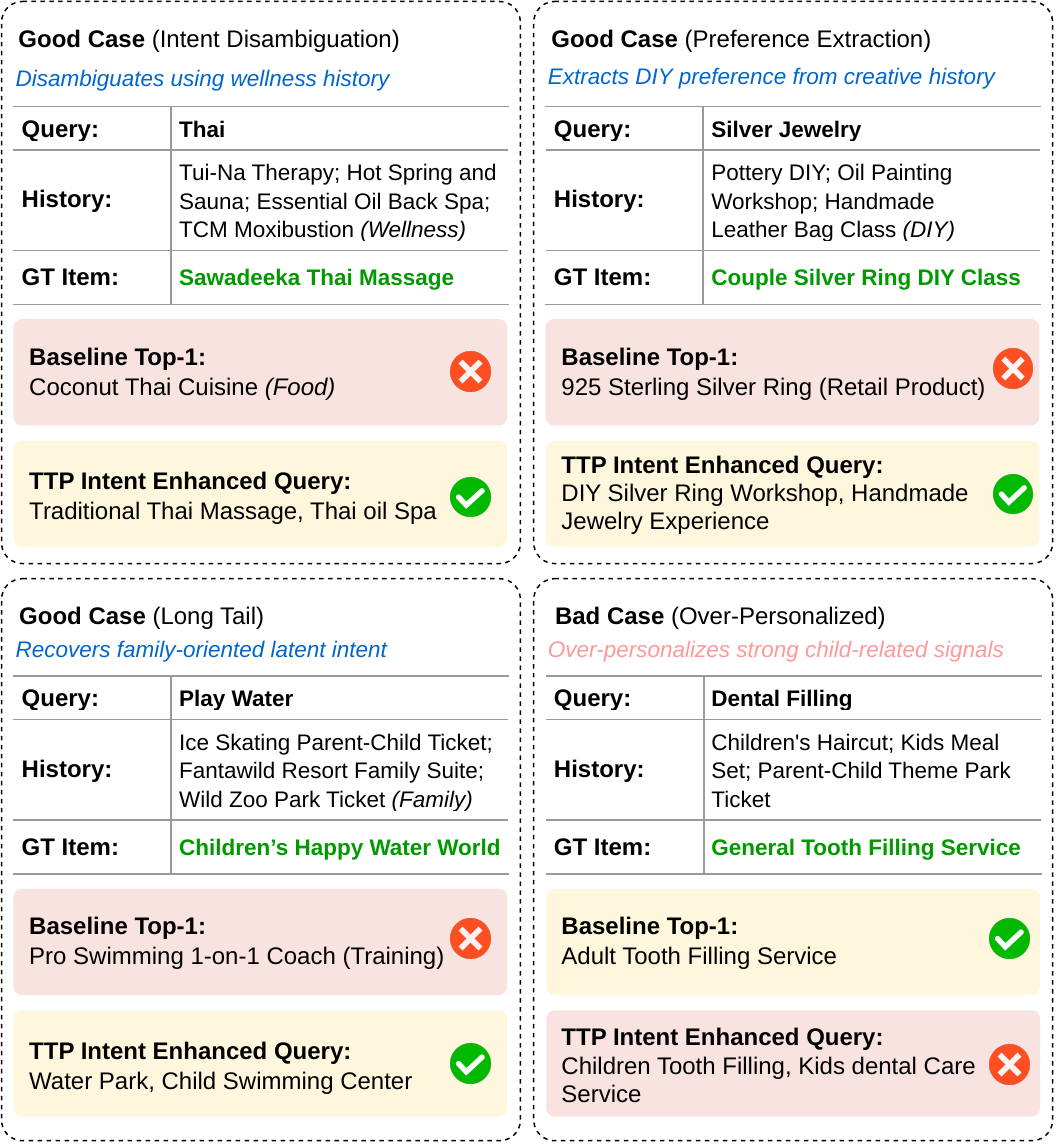}
    \caption{Qualitative case studies of \nameshort{}. \textit{Baseline Top-1} is the top-1 item retrieved by Qwen-Embedding, and \textit{GT Item} is the item finally purchased by the user. For readability, the displayed history only includes representative query-relevant keywords or historical items.}
    \Description{A four-panel qualitative figure. In each panel, the baseline top-1 item is the top-1 result directly retrieved by Qwen-Embedding, and the ground-truth item is the item finally purchased by the user. For readability, the displayed history only contains representative query-relevant keywords or historical items rather than the full user history sequence. The top-left panel shows a good case of intent disambiguation for the query Thai, where wellness-related history helps TTP retrieve a Thai massage service instead of Thai food. The top-right panel shows a good case of preference extraction for the query Silver Jewelry, where DIY-oriented history helps TTP rewrite the query toward a silver ring workshop instead of a retail product. The bottom-left panel shows a good case of long-tail recovery for the query Play Water, where family-entertainment history helps TTP infer a children's water park instead of swimming training. The bottom-right panel shows a bad case of over-personalization for the query Dental Filling, where strong child-related history causes TTP to over-personalize the query into a pediatric dental intent, while the correct target is a general adult dental service.}
    \label{fig:case_study}
\end{figure}

Figure~\ref{fig:rewrite_behavior_distribution} summarizes the rewrite behaviors of \nameshort{}, while Figure~\ref{fig:case_study} provides representative examples. We randomly sample 200 broad-query cases and ask human experts to annotate the generated Intent-Enhanced Queries into five categories: \textit{Keep Original}, \textit{Intent Disambiguation}, \textit{Preference Extraction}, \textit{Query Correction}, and \textit{Over-Personalization}. Among all sampled cases, 34.0\% are kept unchanged. This is expected since not all historical behaviors provide useful evidence for the current query, and preserving the original query is desirable when user history provides no clear query-relevant evidence.

Figure~\ref{fig:rewrite_behavior_distribution} further breaks down the rewritten cases only. Among them, 52.3\% fall into \textit{Intent Disambiguation}, where the original query is ambiguous and user history helps infer the intended meaning, while 40.9\% fall into \textit{Preference Extraction}, where the query intent is already relatively clear and the model further injects personalized preference signals to make the retrieval target more specific. This suggests that when \nameshort{} decides to rewrite, it mainly intervenes in the two most beneficial ways: resolving ambiguous intent and refining the query with personalized preference signals. \textit{Query Correction} accounts for 4.5\% of rewritten cases, while harmful \textit{Over-Personalization} remains limited at 2.3\%, suggesting that the model rarely introduces overly personalized intent.

Figure~\ref{fig:case_study} further illustrates these patterns through three successful cases and one failure case. In each case, \textit{Baseline Top-1} denotes the top-1 item directly retrieved by Qwen-Embedding, while \textit{GT Item} denotes the item finally purchased by the user. For readability, the displayed history only includes representative query-relevant keywords or historical items rather than the user's full behavior sequence. The successful cases show that \nameshort{} can \textbf{(1)} disambiguate broad queries such as ``Thai'' by grounding them in wellness-related history, \textbf{(2)} extract personalized preference signals from queries such as ``Silver Jewelry'' and align them with the user's DIY tendency, and \textbf{(3)} recover family-oriented latent intent for sparse long-tail queries such as ``Play Water''. In contrast, the failure case reveals a limitation of the model: under strong and consistent historical family signals, \nameshort{} may over-generalize past preferences and produce an over-personalized rewrite, such as expanding ``Dental Filling'' into a child-specific dental intent.

\subsection{Ablation Studies}
\subsubsection{Core Component Ablation}

Table~\ref{tab:core_ablation} summarizes the contribution of the core components in \nameshort{}. First, compared with Qwen-Embedding and the stronger Decoupled-Stage baseline, \nameshort{} already achieves clear gains after SFT, demonstrating the benefit of jointly modeling reasoning and retrieval with a shared encoder. Unlike decoupled rewrite-then-retrieve pipelines, this design allows the retrieval objective to directly shape the generated intent representation and reduces the mismatch between query rewriting and retrieval. Second, replacing the Intent-Enhanced Query with the original query during retrieval causes consistent performance drops across all three test sets, confirming that explicit intent reasoning is a key source of improvement rather than an auxiliary generation step. Third, removing Dynamic Positive Selection leads to the largest degradation, showing that filtering noisy rollouts and selecting high-reward candidates is important for stabilizing RL alignment and contrastive learning.

\begin{table}[t]
\centering
\caption{Core component ablation of \nameshort{}.}
\label{tab:core_ablation}
\resizebox{\columnwidth}{!}{
\begin{tabular}{l|cccc}
\toprule
\multirow{2}{*}{\textbf{Method}} & \textbf{General} & \textbf{Broad} & \textbf{LongTail} & \textbf{Avg.} \\
 & \textbf{R@20} & \textbf{R@20} & \textbf{R@20} & \textbf{R@20} \\
\midrule
Qwen-Embedding & 44.20 & 41.27 & 29.76 & 38.41 \\
Decoupled-Stage & 45.89 & 45.08 & 35.18 & 42.05 \\
\nameshort{} (SFT) & 45.38 & 45.66 & 35.70 & 42.25 \\
\nameshort{} (SFT) w/o Intent-Q & 44.48 & 43.68 & 33.66 & 40.61 \\
\nameshort{} (SFT+RL) w/o Intent-Q & 44.81 & 44.93 & 34.30 & 41.35 \\
\nameshort{} (SFT+RL) w/o Dynamic Sele. & 41.39 & 41.02 & 30.71 & 37.71 \\
\textbf{\nameshort{} (SFT+RL)} & \textbf{47.52} & \textbf{48.74} & \textbf{38.65} & \textbf{44.97} \\
\bottomrule
\end{tabular}
}
\end{table}

\begin{figure}[t]
    \centering
    \includegraphics[width=0.85\columnwidth]{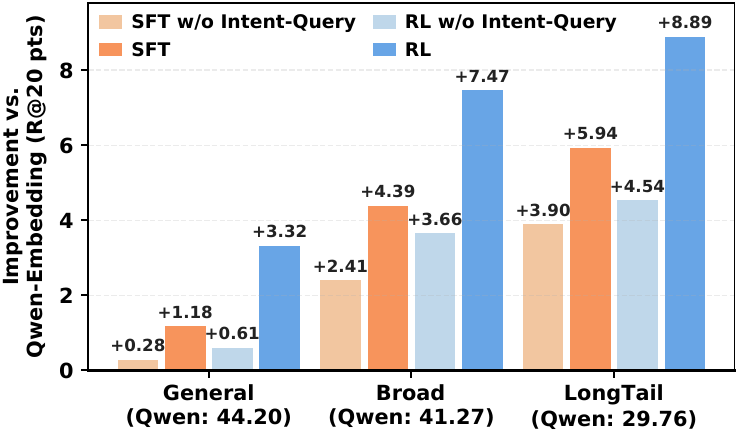}
    \caption{Relative Recall@20 improvement over Qwen-Embedding on the General, Broad, and LongTail test sets. \nameshort{} (SFT) w/o Intent-Q and \nameshort{} (RL) w/o Intent-Q denote the variants where the generated Intent-Enhanced Query is replaced with the original query during retrieval.}
    \Description{A bar chart showing the relative Recall@20 improvement over Qwen-Embedding on the General, Broad, and LongTail test sets for four variants: TTP SFT without Intent-Q, TTP SFT, TTP RL without Intent-Q, and TTP RL. The gains are consistently positive, are larger for the full models than their without-Intent-Q counterparts, and are especially pronounced on the Broad and LongTail sets.}
    \label{fig:ablation_intent}
\end{figure}

\subsubsection{Effect of Intent-Enhanced Query}
Figure~\ref{fig:ablation_intent} reports the relative Recall@20 improvement over Qwen-Embedding on the General, Broad, and LongTail test sets, comparing the full \nameshort{} models with variants that replace the generated Intent-Enhanced Query by the original query during retrieval. Three observations can be drawn from the figure:  
\textbf{(1)} Explicitly generated Intent-Enhanced Queries consistently bring positive gains across both training stages and all three test sets, showing that explicit intent reasoning can effectively bridge the gap between the literal query and the user's latent personalized preference. 
\textbf{(2)} The gain brought by the Intent-Enhanced Query becomes substantially larger after RL than after SFT. For example, on the Broad test set, the additional benefit brought by the Intent-Enhanced Query expands from +1.98\% (=4.39-2.41) to +3.81\% (=7.47-3.66). This larger increment indicates that the RL stage not only improves generation quality, but more importantly aligns the generated intent with downstream retrieval utility. 
\textbf{(3)} Compared with the General set, the gains brought by the Intent-Enhanced Query on Broad and LongTail are consistently larger, suggesting that explicit intent reasoning is especially helpful for ambiguous and sparse queries, where user history provides more critical signals for recovering the actual personalized search intent and thus bridges the intent gap.

\begin{table}[t]
    \centering
    \caption{Ablation study on RL components comparing different selection strategies and reward formulations.}
    \label{tab:ablation_rl}
    \resizebox{\columnwidth}{!}{
    \begin{tabular}{l|ccc|c}
    \toprule
    \multirow{2}{*}{\textbf{Method}} & \textbf{General} & \textbf{Broad} & \textbf{LongTail} & \textbf{Avg.} \\
     & \textbf{R@20} & \textbf{R@20} & \textbf{R@20} & \textbf{R@20} \\ \midrule
    \nameshort{} (SFT Only) & 45.38 & 45.66 & 35.70 & 42.25 \\ \midrule
    \multicolumn{5}{l}{\textit{\nameshort{} (SFT+RL)}} \\
    \quad w/o Dynamic Selection & 41.39 & 41.02 & 30.71 & 37.71 \\
    \quad w/ Unfrozen Reward Model & 32.48 & 33.74 & 22.36 & 29.53 \\
    \quad w/o Positive Gain ($\alpha$) & 45.18 & 45.75 & 35.29 & 42.07 \\
    \quad w/o Margin Gain ($\beta$) & \underline{46.82} & \underline{47.16} & \underline{37.73} & \underline{43.90} \\ \midrule
    \textbf{\nameshort{} (SFT+RL)} & \textbf{47.52} & \textbf{48.74} & \textbf{38.65} & \textbf{44.97} \\ \bottomrule
    \end{tabular}
    }
    \end{table}

    \begin{figure}[t]
        \centering
        \captionsetup[subfigure]{font=scriptsize}
        \begin{subfigure}[b]{0.24\columnwidth}
            \centering
            \includegraphics[width=\textwidth]{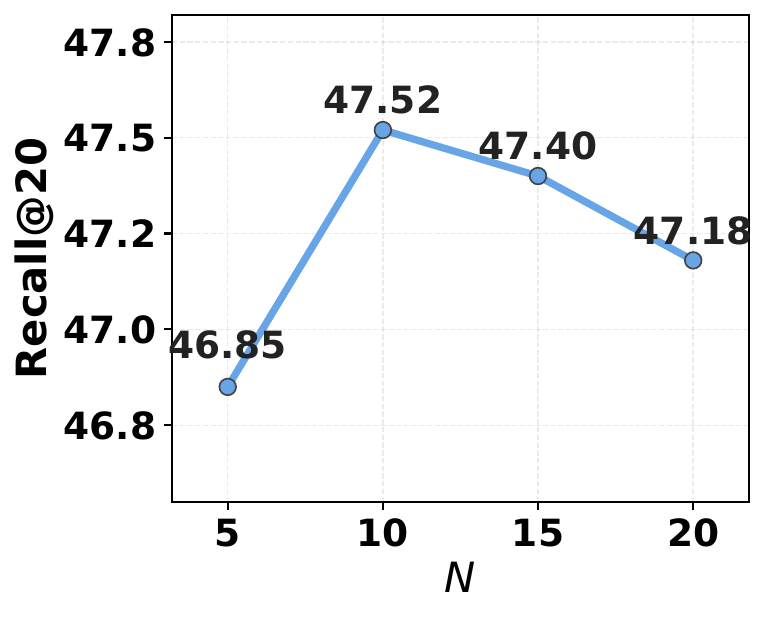}
            \caption{General}
            \label{fig:hist_gen}
        \end{subfigure}
        \hfill
        \begin{subfigure}[b]{0.24\columnwidth}
            \centering
            \includegraphics[width=\textwidth]{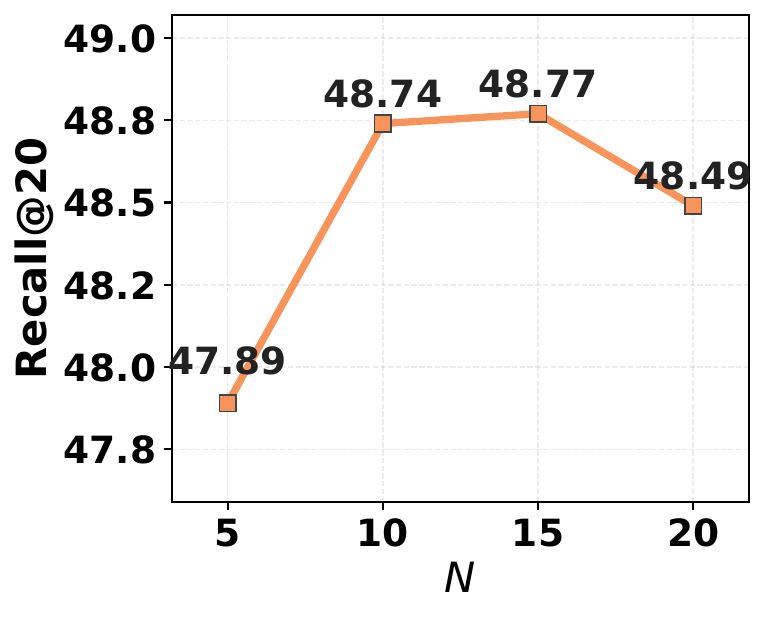}
            \caption{Broad}
            \label{fig:hist_broad}
        \end{subfigure}
        \hfill
        \begin{subfigure}[b]{0.24\columnwidth}
            \centering
            \includegraphics[width=\textwidth]{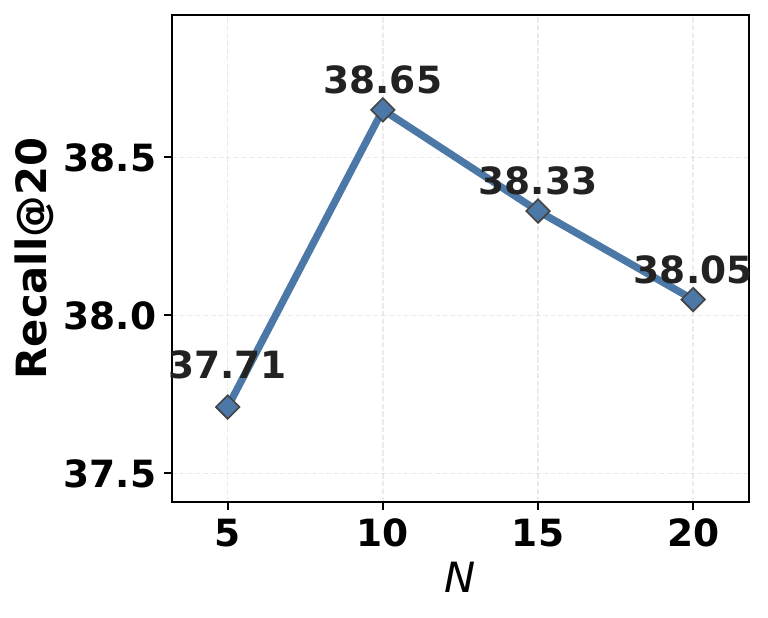}
            \caption{LongTail}
            \label{fig:hist_long}
        \end{subfigure}
        \hfill
        \begin{subfigure}[b]{0.24\columnwidth}
            \centering
            \includegraphics[width=\textwidth]{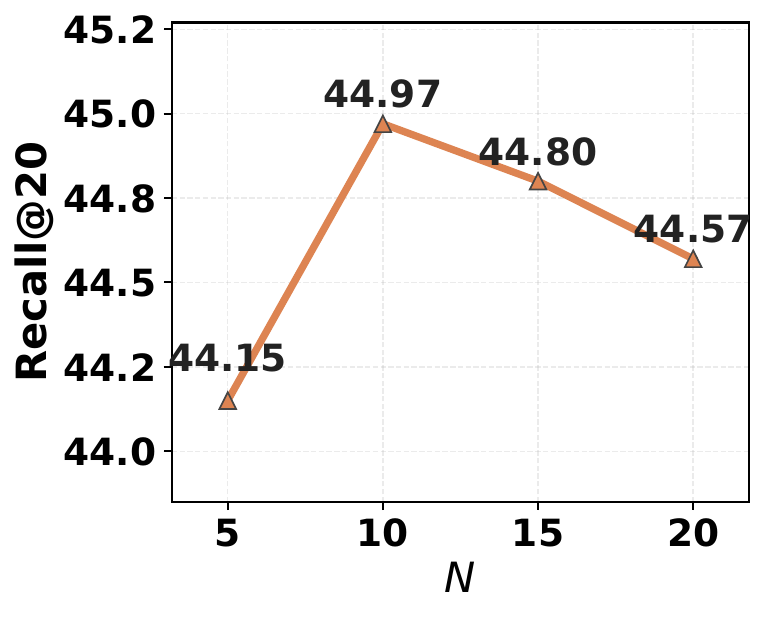}
            \caption{Average}
            \label{fig:hist_avg}
        \end{subfigure}
        
        \caption{Impact of history sequence length on retrieval performance (Recall@20).}
        \Description{Four line charts illustrating the impact of history sequence max length (x-axis: 5, 10, 15, 20) on retrieval performance (y-axis: Recall@20) across three datasets: General, Broad, Longtail, together with their average. In all four charts, the performance initially increases as the history length grows from 5 to 10, reaching a peak at length 10. As the length increases further to 15 and 20, the performance shows a slight decline in all scenarios, indicating that a history length of 10 is optimal.}
        \label{fig:effect_history_length}
    \end{figure}

\subsubsection{Effect of RL Training Strategies}
Table~\ref{tab:ablation_rl} analyzes the impact of different components in the RL stage. We compare the full model against four variants: (1) \textit{w/o Dynamic Selection}, which uses random selection from the group as the positive query for contrastive learning; (2) \textit{w/ Unfrozen Reward Model}, where the reward signal is computed directly by the training policy model itself rather than the frozen reference SFT model; and (3) \textit{w/o Positive Gain ($\alpha$)} and \textit{w/o Margin Gain ($\beta$)}, which remove the absolute score gain and margin gain terms from the retrieval reward, respectively.

As shown in Table~\ref{tab:ablation_rl}, using random selection (\textbf{\textit{w/o Dynamic Selection}}) exhibits a significant performance drop (Average Recall@20: 37.71\%), worse than the SFT baseline (42.25\%). This indicates that the rollout process in GRPO frequently generates noisy or low-quality queries, and randomly selecting these as positive samples for contrastive learning introduces severe noise. In contrast, dynamic positive selection filters out low-quality queries using the reward signal of RL, bridging the gap between generation and retrieval, leading to better end-to-end alignment of the generated intent with the retrieval task.

Notably, allowing the training policy model to evaluate itself (\textbf{\textit{w/ Unfrozen Reward Model}}) results in a catastrophic performance drop to 29.53\%. This confirms that using the unfrozen actor as its own judge suffers from severe self-referential bias, causing the model to exploit its own scoring artifacts (reward hacking) rather than optimizing for true relevance. By contrast, the frozen SFT model serves as a stable semantic anchor that is intrinsically aligned with the final retrieval objective.
This stability is critical for effective alignment, providing consistent supervision that guides the policy to enhance its reasoning capabilities without drifting into unstable semantic spaces.

Removing the positive gain term (\textbf{\textit{w/o Positive Gain}}) leads to a substantial drop to 42.07\%, while removing the margin gain term (\textit{w/o Margin Gain}) results in a smaller decrease to 43.90\%. This suggests that both positive gain and margin gain contribute to the final performance. The positive gain term encourages the model to directly improve the relevance over the original query, while the margin gain term helps to further distinguish the positive item from the negative items.

\subsubsection{Effect of History Sequence Length}
We vary the maximum history length $N \in \{5, 10, 15, 20\}$ during training and inference, selecting the top-$N$ items from the full history with \textit{bge-reranker-v2-m3}. Figure~\ref{fig:effect_history_length} reports Recall@20 on General, Broad, and LongTail. Performance peaks at $N=10$; longer histories introduce noise and yield diminishing returns. We therefore set $N=10$ to balance historical context and efficiency.

\section{Deployment and Online Results}

\subsection{Online Deployment}
\begin{figure}[t]
    \centering
    \includegraphics[width=\columnwidth]{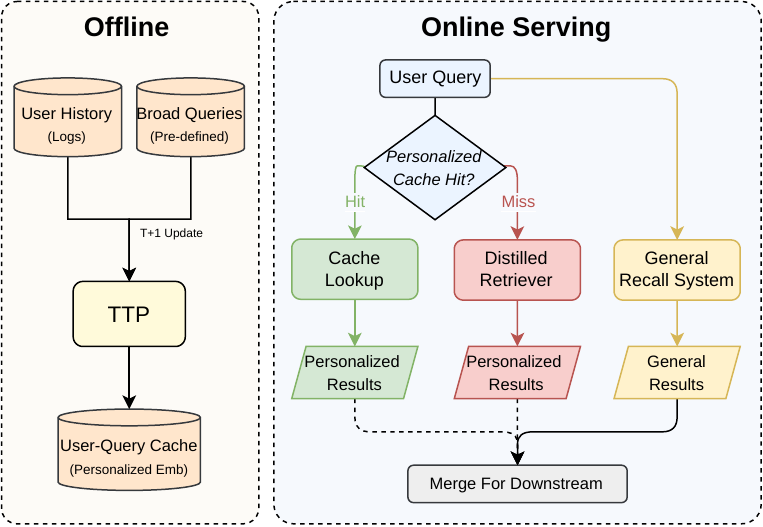}
    \caption{Hybrid online serving framework of \nameshort{}.}
    \Description{A system diagram of the hybrid online serving framework. Offline, the model builds a personalized cache from user history logs and a broad-query set, and pre-computes item embeddings. Online, a user query first checks the personalized cache. If it hits, the system performs embedding cache lookup and ANN search to produce personalized recall results. If it misses, the system uses a distilled online model to produce personalized recall results. In parallel, a general recall system produces general recall results. The personalized and general candidates are then merged and sent to downstream processing.}
    \label{fig:online_deploy}
\end{figure}

To support efficient online serving, we develop a hybrid framework that combines an offline personalized cache with a lightweight distilled retriever, as illustrated in Figure~\ref{fig:online_deploy}.

\paragraph{Offline Personalized Cache.}
Through T+1 daily updates, the full LLM-based \nameshort{} model pre-computes intent-enhanced query embeddings for active user--broad-query pairs. For cache hits, the system directly performs ANN search using the stored embeddings. This cache-hit path covers 10.59\% of query volume (QV) and 11.80\% of user volume (UV), while cache misses are handled by the distilled retriever.

\paragraph{Distilled Real-Time Retrieval.}
For cache misses, we distill the personalized rewriting capability of \nameshort{} into an encoder-only bi-encoder using InfoNCE~\cite{xia2026reasoning}. We construct an original-query view $q$ and an intent-enhanced view $q_r$ with the same \texttt{instruction + history + query} template, and denote the positive item by $p^+$. The model jointly optimizes $\langle q,p^+\rangle$, $\langle q_r,p^+\rangle$, and $\langle q,q_r\rangle$ with in-batch negatives and equal weights (1:1:1). As shown in Table~\ref{tab:distill_results}, the latter two objectives transfer personalized rewrite semantics into the original-query representation while retaining most of the full model's retrieval quality.

\begin{table}[t]
    \centering
    \caption{Ablation study of the distilled retriever.}
    \label{tab:distill_results}
    \resizebox{0.9\columnwidth}{!}{
    \begin{tabular}{l|cccc}
    \toprule
    \multirow{2}{*}{\textbf{Method}} & \textbf{General} & \textbf{Broad} & \textbf{LongTail} & \textbf{Avg.} \\
     & \textbf{R@20} & \textbf{R@20} & \textbf{R@20} & \textbf{R@20} \\ \midrule
    Full \nameshort{} (Teacher, 3B) & \textbf{47.52} & \textbf{48.74} & \textbf{38.65} & \textbf{44.97} \\ \midrule
    Distilled Retriever (305M) & \underline{46.14} & \underline{47.45} & {35.72} & \underline{43.10} \\
    \quad w/o $\langle q_r, p^+ \rangle$ & 45.59 & 46.56 & \underline{35.88} & 42.68 \\
    \quad w/o $\langle q, q_r \rangle$ & 44.98 & 44.73 & 34.19 & 41.30 \\
    \quad w/o $\langle q_r, p^+ \rangle$ and $\langle q, q_r \rangle$ & 44.67 & 44.21 & 33.26 & 40.71 \\
    \bottomrule
    \end{tabular}
    }
\end{table}

\begin{figure}[t]
    \centering
    \includegraphics[width=0.9\columnwidth]{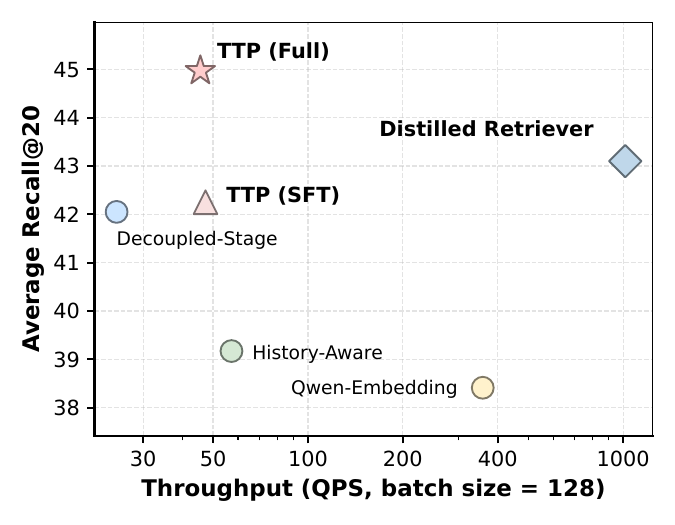}
    \caption{Throughput--performance trade-off on one NVIDIA A100 80GB GPU (batch size 128), measured by average Recall@20 across General, Broad, and LongTail.}
    \Description{A scatter plot comparing different retrieval architectures in terms of throughput and performance. The x-axis is throughput in QPS measured on one NVIDIA A100 80GB GPU with batch size 128. The y-axis is average Recall at 20 across the General, Broad, and LongTail datasets. The plot includes methods such as Qwen-Embedding, History-Aware, a decoupled two-stage method, TTP SFT, TTP Full, and a distilled model.}
    \label{fig:throughput_performance}
\end{figure}

\paragraph{Serving Pipeline and Efficiency.}
The personalized channel runs in parallel with the existing general recall system. Their candidates are merged under a fixed budget and sent to the same downstream ranking pipeline. As shown in Figure~\ref{fig:throughput_performance}, the full \nameshort{} model is more efficient than the decoupled two-stage pipeline, while the distilled retriever achieves a favorable effectiveness-efficiency trade-off for real-time serving.

\subsection{Online A/B Test}

\begin{table}[t]
    \centering
    \caption{Seven-day online A/B test results (relative lift). All effectiveness gains are significant at $p<0.05$; latency denotes additional end-to-end overhead.}
    \label{tab:ab_test}
    \resizebox{0.45\columnwidth}{!}{
    \begin{tabular}{l|c}
    \toprule
    \textbf{Metric} & \textbf{Relative Lift} \\
    \midrule
    QV CXR & +0.41\% \\
    UV CXR & +0.29\% \\
    Order Volume & +0.46\% \\
    \midrule
    Avg. Latency & +0.15 ms \\
    \bottomrule
    \end{tabular}
    }
\end{table}

We conduct a 7-day online A/B test by introducing \nameshort{} as an additional personalized recall channel while keeping the general recall system, candidate budget, and downstream ranking pipeline unchanged. As shown in Table~\ref{tab:ab_test}, \nameshort{} improves QV CXR, UV CXR, and order volume by 0.41\%, 0.29\%, and 0.46\%, respectively, with only 0.15 ms additional end-to-end latency. The low overhead is enabled by the cache/distillation design and parallel execution with general recall. The system has been deployed on our platform.

%% file: sections/conclusion.tex
\section{Conclusion and Future Work}
\label{sec:conclusion}

We presented \nameshort{}, a unified framework that advances personalized dense retrieval from implicit behavior modeling to explicit intent reasoning. Through SFT and retrieval-aligned RL, \nameshort{} jointly optimizes personalized intent generation and dense retrieval, effectively bridging the intent gap for ambiguous and long-tail queries. Extensive evaluations on proprietary and public benchmarks, together with its large-scale online deployment, demonstrate the effectiveness and practical value of our approach. Future work will explore multi-turn personalization and efficient real-time inference.